# THE IMPACT OF LEADERSHIP STYLES ON PROJECT EFFICIENCY

Michał ĆWIĄKAŁA[1*], Julia WALTER[2], Dariusz BARAN[3], Gabriela WOJAK[4], Ernest GÓRKA[5], Piotr MRZYGŁÓD[6], Maciej FRASUNKIEWICZ[7], Piotr RĘCZAJSKI[8], Jan PIWNIK[9]

[1] I'M Brand Institute sp. z o.o.; m.cwiakala@imbrandinstitute.pl, ORCID: 0000-0001-9706-864X
[2] I'M Brand Institute sp. z o.o.; j.walter@imbrandinstitute.pl, ORCID: 0009-0005-2979-1170
[3] WSB - National-Louis University, College of Social and Computer Sciences; dkbaran@wsb-nlu.edu.pl, ORCID: 0009-0006-8697-5459
[4] Jagiellonian University, Faculty of Management and Social Communication; gabriela.wojak@student.uj.edu.pl, ORCID: 0009-0003-2958-365X
[5] WSB - National-Louis University, College of Social and Computer Sciences; ewgorka@wsb-nlu.edu.pl, ORCID: 0009-0006-3293-5670
[6] Piotr Mrzygłód Sprzedaż-Marketing-Consulting; piotr@marketing-sprzedaz.pl, ORCID: 0009-0006-5269-0359
[7] F3-TFS sp. z o.o.; m.frasunkiewicz@imbrandinstitute.pl, ORCID: 0009-0006-6079-4924
[8] MAMASTUDIO Pawlik, Ręczajski, spółka jawna; piotr@mamastudio.pl, ORCID: 0009-0000-4745-5940
[9] WSB Merito University in Gdańsk, Faculty of Computer Science and New Technologies; jpiwnik@wsb.gda.pl, ORCID: 0000-0001-9436-7142
* Correspondence author

**Purpose:** This study aims to investigate the impact of various leadership styles on project efficiency within different organizational contexts. It seeks to identify the key leadership behaviors that contribute to the successful attainment of project goals, stakeholder satisfaction, and overall project performance.
**Design/methodology/approach**: Quantitative research using a questionnaire with 100 project professionals. Data analyzed through statistical methods, including Spearman's correlation.
**Findings:** Leadership styles significantly affect project efficiency. Constructive feedback, clear goals, role clarity, and team initiative are key factors.
**Research limitations/implications**: Limited sample size and scope. Future studies should include diverse industries and cultural contexts.
**Practical implications:** Highlights the need for clear communication, feedback, and team empowerment to improve project outcomes.
**Social implications:** Effective leadership fosters trust, collaboration, and employee satisfaction, boosting productivity.
**Originality/value:** Provides empirical evidence on leadership styles' effect on project efficiency, offering actionable insights for managers.
**Category of the paper:** research paper.

## 1. Introduction

This study provides an original contribution by empirically exploring the nuanced relationship between leadership styles and project efficiency, a topic often discussed theoretically but seldom quantified in practice. By using a unique dataset gathered directly from project participants across multiple industries, it offers fresh insights and practical guidance that bridge the existing gap between abstract leadership concepts and measurable project outcomes. Additionally, this paper provides empirical verification of specific leadership behaviors, offering quantitative evidence of their direct influence on project success, thus enriching existing theoretical discussions.

Leadership plays a pivotal role in determining the success and efficiency of project implementation within various organizational contexts. This study was undertaken to investigate the extent to which different leadership styles influence project outcomes, particularly in terms of efficiency, goal attainment, and stakeholder satisfaction. The primary research question guiding this study is: *How do different leadership styles impact the efficiency of project teams?*

The theoretical framework of this research is grounded in classical and contemporary leadership theories. It encompasses foundational models such as trait theory, behavioral theory, situational leadership, and transformational leadership, extending to modern approaches like servant leadership and participative leadership. These theories provide the basis for understanding the diverse mechanisms through which leaders can affect team dynamics, decision-making processes, and overall project performance.

The motivation for this study arises from the recognition of a significant knowledge gap in understanding the direct correlation between specific leadership styles and measurable project efficiency outcomes. While extensive literature exists on leadership theories and their qualitative impact on organizations, there is limited empirical evidence quantifying these effects in project-based environments. This gap necessitates a focused investigation to provide actionable insights for practitioners and scholars alike.

This study aims to identify and categorize prevalent leadership styles within project teams, analyze the relationship between these leadership styles and key project performance indicators, and determine the leadership behaviors most conducive to enhancing project efficiency. The study employs a quantitative research methodology, utilizing a structured questionnaire distributed to professionals working in project teams across various industries. The collected data is analyzed using statistical techniques to establish correlations between leadership styles and project efficiency metrics.

By addressing the outlined research question and objectives, this study aims to contribute to the academic discourse on leadership in project management and offer practical recommendations for optimizing leadership practices to improve project outcomes.

## 2. The beginnings of leadership theories

Leadership concepts have evolved over the years. Initially, traditional approaches to leadership, which focused on leaders' personality traits, transformed into more complex and flexible relational and situational models. The development process that the concept of leadership has undergone, as well as the theories concerning it, influences how leadership is perceived today. Therefore, it is worthwhile to recount the history of theories formulated about leaders, their traits, and actions to gain the perspective necessary for analyzing contemporary leadership models.

One of the earliest theories aimed at explaining the essence of effective leadership was the trait theory. This theory suggested that leaders are individuals who inherently possess qualities like intelligence, charisma, communication skills, determination, and self-confidence (Northouse, 2016). In Polish literature, Tadeusz Kotarbiński contributed to the development of trait theory by emphasizing the critical importance of moral values and the responsibility leaders bear in their work (Kotarbiński, 2000). While this theory enjoyed significant popularity in the first half of the 20th century, its prominence declined as certain limitations became evident. Observations indicated that not all people exhibiting these traits become effective leaders. It was recognized that innate character traits and ethical values, though important, are not the sole determinants of a leader's success. Additionally, it was noted that individuals in leadership roles can develop their competencies and traits over the course of their careers. This realization suggested that the trait theory was too simplistic to fully explain what characterizes a successful leader (Zaccaro, 2007).

Further advancements in leadership research emerged in the early part of the second half of the 20th century. The focus shifted from examining innate traits of leaders to analyzing their behaviors. This shift led to the development of behavioral theories, which posited that a leader's actions are the primary factors influencing their effectiveness (Yukl, 2013). However, these theories also had limitations and were not adequate for explaining all situations.

Paul Hersey and Kenneth Blanchard developed the situational leadership theory to address the shortcomings identified in earlier leadership theories. At the time of its creation, this concept was considered highly innovative. It builds upon previous observations and is based on the belief that an effective leader adapts their approach to match the readiness level of their subordinates. The leader selects their style of behavior to appropriately guide the project and fulfill its requirements (Yukl, 2013; Blanchard, 2007). The emergence of this theory significantly impacted the progression of other contemporary project management methods by highlighting the importance of leader flexibility when managing a project. Research in this area has shown that this management style is particularly effective in projects characterized by dynamic changes in economic conditions (Koźmiński, Jemielniak, 2011).

Another leadership approach that gained popularity in the second half of the 20th century is the relationship-oriented management style. As the name suggests, this style focuses on building strong relationships between the leader and team members. Leaders who employ this style make an effort to understand their coworkers' needs, support their development, and maintain mutual respect and trust within the team (Bass, Riggio, 2006). Obłój identified that fostering relationships with the team positively influences the project's final results, as team members become engaged and feel responsible for the tasks entrusted to them. Therefore, this concept is particularly effective in project teams where the nature of the tasks requires close cooperation and high-level communication (Obłój, 2007).

Classical leadership concepts have undeniably shaped how contemporary models are utilized today. The theories previously discussed have been updated with continually expanding knowledge in this field and adjusted to align with current realities. They have provided a robust foundation for the development of management models such as transformational leadership and servant leadership (Yukl, 2013). Understanding and analyzing classical leadership theories enables leaders to better adapt modern models in their work. This also facilitates the development, enhancement, and customization of practices within modern project teams (Koźmiński, Jemielniak, 2011).

## 3. Contemporary leadership styles

The evolution of modern leadership models arose from the need for more complex and flexible leadership techniques. Although contemporary approaches have roots in classical leadership concepts, they have been expanded and adapted to suit changing times and working conditions. We include among contemporary the following leadership styles:
- Transformational leadership (Bass, Riggio, 2006),
- Transactional leadership (Burns, 1978),
- Participative leadership (Stocki et al., 2008),
- Democratic leadership (Gastil, 1994),
- Servant leadership (Greenleaf, 1977),
- Situational leadership (Blanchard, 2007).

Transformational leadership centers on inspiring teams to achieve shared goals through engagement and vision. Leaders who adopt this style focus on motivating employees, enhancing their commitment, and developing their personal competencies (Bass, Riggio, 2006). This approach emphasizes creating strong emotional bonds within the team. Key characteristics include the leader's charisma, individualized consideration for team members, and actions aimed at intellectually stimulating employees. Teams are encouraged to think creatively and innovate (Chmiel, 2003). In practical terms, transformational leadership is particularly significant in projects requiring high levels of innovation. Leaders can inspire colleagues to seek new solutions continuously, positively impacting project outcomes (Bass, Riggio, 2006).

Conversely, transactional leadership is based on a clear system of rewards and penalties—a transactional relationship between the leader and the team. This approach assumes that employees are most motivated and perform at their best when leaders establish clearly defined goals and provide appropriate rewards (material or non-material) for their achievement. Inadequate results may lead to consequences for employees (Burns, 1978). This style is most effective in projects requiring strict control and precision, such as those with high operational complexity and significant risk. Essentially, transactional leadership focuses on motivating employees through a straightforward system of incentives and sanctions. (Schultz, Schultz, 2006).

Despite their differences, these two leadership styles are often used together to complement each other. Combining them is effective in teams where leaders need to manage both creative and operational aspects efficiently. By drawing from the transformational style, leaders focus on inspiring the team to achieve visionary goals by stimulating innovative solutions and adapting to new challenges. Utilizing the transactional approach ensures that daily work aligns with project assumptions, taking care of task execution within the set time and budget (Bass, Avolio, 1994). Employing a hybrid of these styles is effective in complex projects requiring flexibility, creativity, and precise operational management. Leaders guiding their teams in this manner balance fostering innovation with maintaining high efficiency levels (Avolio, Bass, 2004).

Participative leadership is a management approach in which organizational members actively participate in decision-making, having genuine influence over setting goals and determining the company's direction of development. This method enhances employee engagement because they feel valued and partly accountable for the organization's success. Collaboration under this model encourages the growth of personal competencies within the team, contributing to overall organizational advancement. Adopting participative leadership involves decentralizing authority, which requires the leader to be open-minded and trusting toward employees (Stocki et al., 2008). In project teams, this style increases effectiveness, especially in settings that demand innovative solutions to problems.

Similarly, democratic leadership entails equal involvement of all team members in decision-making processes. This style is ideal for teams that need a facilitator or advisor rather than a traditional manager. Team members develop a sense of responsibility for the success and efficiency of their work. Implementing democratic leadership necessitates strong collaboration skills from all participants, fostering creativity and enabling the full use of the team's collective knowledge and abilities. However, it may introduce challenges in time management since reaching decisions can take longer compared to authoritarian styles. In rapidly changing project environments where swift decision-making is crucial, this can be particularly challenging (Gastil, 1994).

The concept of servant leadership is based on the belief that a leader should act as a servant. It means prioritizing the needs of coworkers and supporting their personal development rather than directing the team authoritatively. Introduced by Robert K. Greenleaf (1977), a servant leader operates with humility and focuses on the growth and needs of team members. Fundamental features of this concept include the leader's modesty and lack of desire to exert power over employees, active listening to understand coworkers' needs better, and a commitment to helping others achieve their full potential rather than focusing on personal career advancement (Blanchard, 2007). It also promotes building interpersonal relationships and fostering an atmosphere of shared responsibility (Spears, 2010). Empathy is essential for building mutual trust and improving team communication (Spears, 1995).

In servant leadership, the leader embodies the principles and values underlying the concept. Initially, they play a visionary or strategic role, conveying the organization's mission and values and outlining general goals and directions. Once employees have the necessary knowledge, the leader's role shifts to being a servant, and the organizational hierarchy transforms. Employees then take responsibility for meeting customer expectations, while the leader responds to their needs (Wlizło, 2021). In project work, adopting servant leadership facilitates greater team autonomy, effective communication, and team development through the personal growth of its members (Blanchard, 2007).

Although previously described as a classical management style due to its innovativeness at the time of inception, situational leadership remains relevant and widely used still today. It is based on flexibility and the ability to adapt actions to specific situations. The leader adjusts their style considering the team's competence, commitment, and maturity. Typically, a directive style at the beginning gradually evolves into a delegating and supporting style (Blanchard, 2007). This approach places significant emphasis on the leader's abilities. Blanchard, Zigarmi, and Nelson (1993) reviewed situational leadership 25 years after its introduction, highlighting its effectiveness and popularity. They concluded that leaders using this concept must continually enhance their skills, especially in assessing and adapting to team needs. Essential leader competencies include flexibility, accurate assessment of employees' readiness and commitment levels, open and clear communication, delegation of responsibility as the team matures, and the ability to motivate employees by tailoring strategies to their needs.

While flexibility and adaptability are core to situational leadership, technological progress and globalization have added new challenges. Leaders frequently oversee global teams dispersed across various regions, requiring them to navigate time zone differences, cultural diversity, and distinct communication styles. This scenario demands exceptional flexibility and the ability to manage diversity effectively (Northouse, 2016). In project management, situational leadership is highly effective because it allows leaders to adjust their approach to both tasks and team members in a flexible manner. This style is particularly advantageous in dynamic projects where requirements, goals, or external factors may change during the project's lifecycle (Northouse, 2016). Additionally, agile project management methodologies incorporate many values from situational leadership. Similarities can be observed between the approaches of Scrum Masters and situational leaders (Rubin, 2012). Many of these leadership styles are widely used today, as they provide effective strategies for various organizational challenges. Leaders often combine elements from different styles to tailor their approach to their team's specific needs.

As Bwalya (2023) described, leadership style plays a significant role in shaping the success and effectiveness of a leader in various organizational contexts. As highlighted in recent literature, there is no universal or optimal leadership style suitable for all situations. Instead, effective leadership is highly dependent on the leader's ability to assess the context and adapt accordingly. Each style has distinct strengths, limitations, and practical implications. A deeper understanding of these elements enables leaders to maximize their impact by aligning their behaviors and strategies with team needs and organizational goals. According to Bwalya (2023), leadership style embodies a leader's values, attitudes, and behaviors and strongly influences team motivation, communication patterns, and workplace culture. Leaders must consciously select and adapt their styles to enhance team dynamics, productivity, and long-term performance. The ability to do so not only promote trust and engagement but also fosters a resilient and adaptive organizational climate.

Employee engagement has become a critical factor for organizational success, as it directly influences productivity, commitment, and workplace culture. According to Shrivastava and Mathur (2025), there is a clear and significant relationship between leadership style and employee engagement. Leaders who adopt a relationship-oriented approach—showing care, support, and commitment to employee development - positively influence engagement levels. Furthermore, a collaborative work environment, where coworkers are perceived as allies rather than competitors, reinforces this effect. Clear and reasonable job expectations also contribute to higher engagement by reducing ambiguity and stress.

## 4. Leadership styles and research methodology

The authors conducted their own survey-based research using a questionnaire. A quantitative method was chosen, which allows for the collection of a large amount of data and the conduction of statistical analysis.

The research tool is a questionnaire constructed using Google Forms, which was distributed to respondents working in project teams. The survey was completed by 100 individuals: 50 women, 46 men, and 4 individuals who preferred not to disclose their gender. The respondents represented various industries including IT and technology, public sector and administration, and marketing and advertising. The questionnaire began with questions on socio-demographic characteristics (gender, age, working status and country). The questionnaire consists of 16 closed questions with two or three answers. The format of the questions and answers in the questionnaire was designed to be as accessible as possible for the respondents. Below is a template of the questionnaire with questions and possible answers.

The aim of this study is to understand the impact of different leadership styles on the effectiveness of project teams. To answer the research question, a quantitative approach was employed, based on questionnaires distributed to professionals working in project teams across various industries. The collected data were analyzed using statistical techniques to identify correlations between leadership styles and project efficiency metrics.

One of the main limitations of the study is the subjectivity of the responses obtained from the participants. When evaluating leadership styles, respondents may be influenced by personal preferences or professional experiences, which could affect the results of the study. Additionally, the study, relying solely on questionnaires, may not capture the full range of variables influencing project effectiveness, such as resources, organizational structure, or industry context. Due to the cross-sectional nature of the study, its results may not reflect the long-term impact of leadership styles on project outcomes, nor account for changes occurring in project teams over time.

The survey data were analyzed using statistical and correlation methods. After data collection, results were visualized using bar charts and pie charts to highlight key trends and distributions. Bar charts were employed to compare specific variables across different categories, while pie charts illustrated proportional relationships within the datasets. Additionally, stacked bar charts, grouped bar charts, and ring charts were utilized for more detailed visual representations. The data from one section of the survey were organized into tables, with all visualizations and calculations performed in Microsoft Excel.

Statistical analysis focused on calculating percentile distributions for all responses. Mean and median values were computed for certain datasets to rank evaluated elements from best to worst. Responses were scored using a scale where "1" represented the most favorable answer, "0.5" was neutral, and "0" was the least favorable. These statistical metrics provided a comprehensive understanding of central tendencies and patterns in the data.

In the correlation analysis, the relationship between constructive feedback from leaders and achieving project goals was examined using Spearman's rank correlation. Constructive feedback from leaders was treated as the independent variable, coded as "Yes" = 1, "Varies" = 0.5, and "No" = 0. Achieving project goals was the dependent variable, coded as "Yes" = 1 and "No" = 0. With a sample size of 100 respondents, the analysis determined the strength and direction of the relationship between these variables. Spearman's method was chosen due to the ordinal nature of the data. The significance of the correlation was also tested to verify statistical reliability, ensuring a robust interpretation of the results. All calculations were conducted using Microsoft Excel, enabling clear comparisons and identifying patterns within the observed relationships.

### 4.1. Findings

A collective summary of the research results for all 100 respondents, in accordance with the 16 questions in the survey, is presented in Figures 1 to 9.

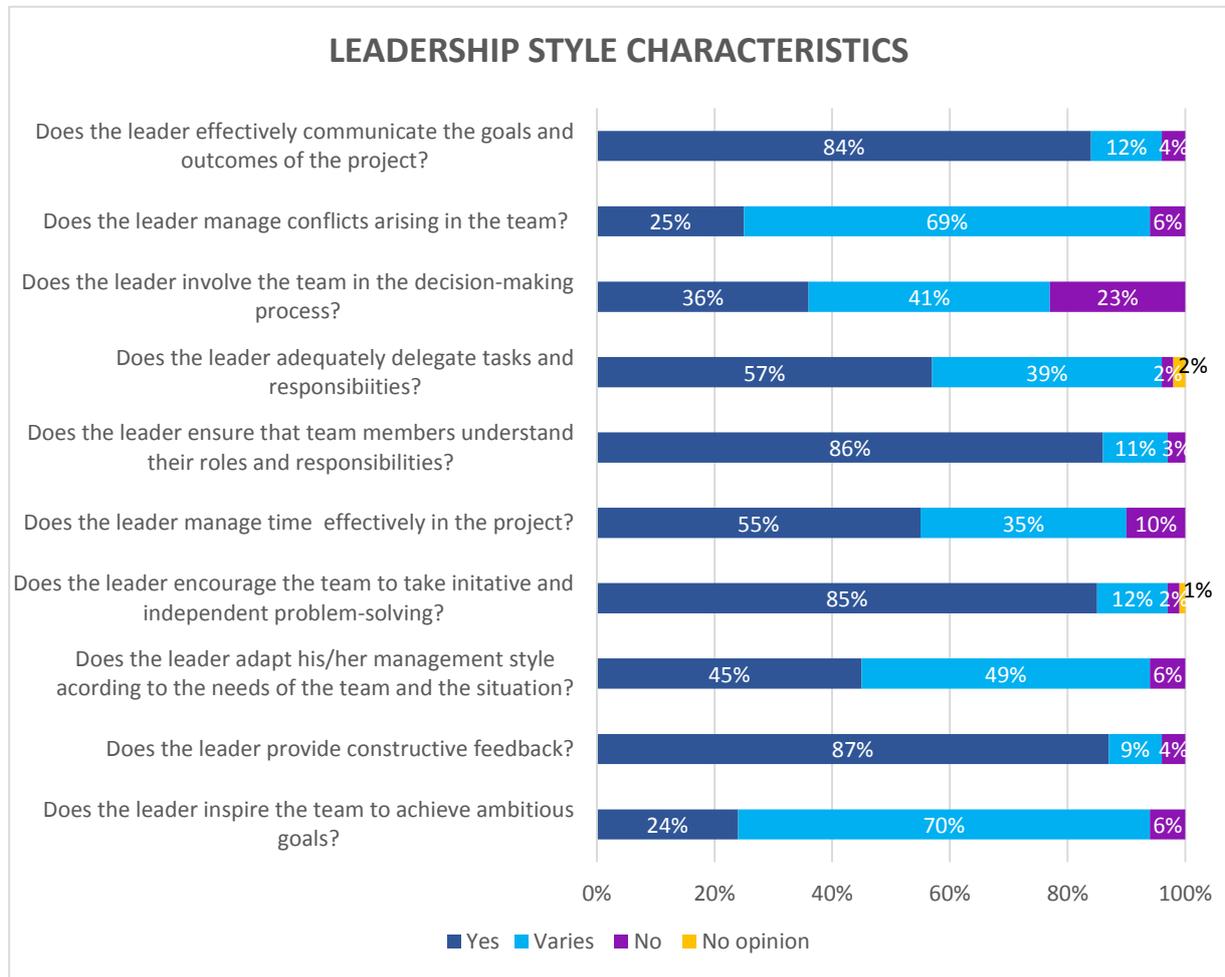

**Figure 1.** Answers from Project Team Members of enterprises from different sectors to questions 1-10 of the research survey.

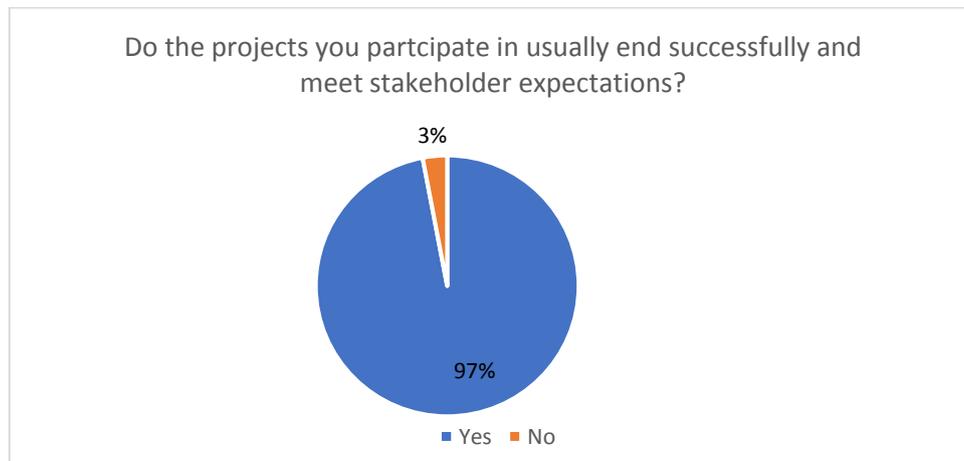

**Figure 2.** Answers from Project Team Members of enterprises from different sectors to question 11 of the research survey.

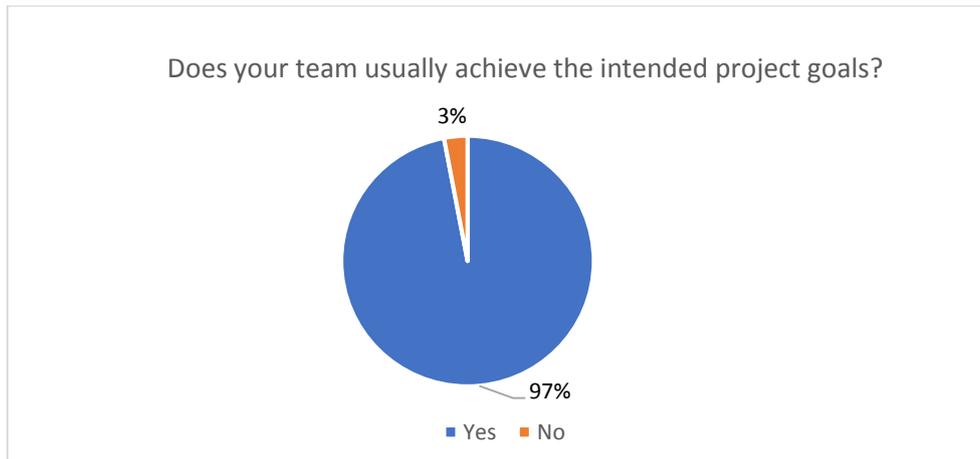

**Figure 3.** Answers from Project Team Members of enterprises from different sectors to question 12 of the research survey.

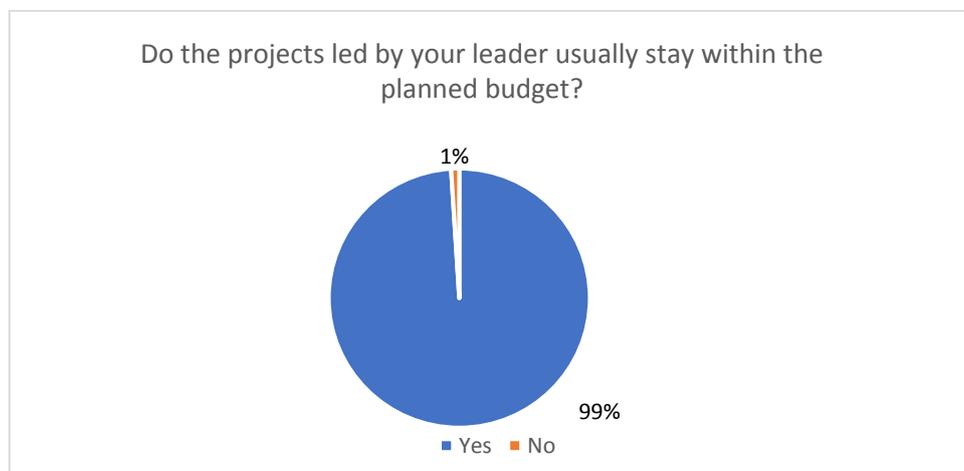

**Figure 4.** Answers from Project Team Members of enterprises from different sectors to question 13 of the research survey.

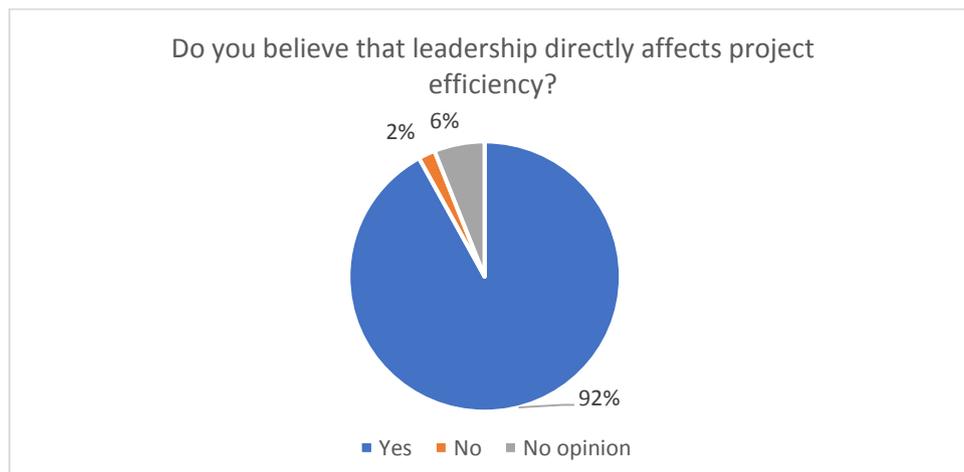

**Figure 5.** Answers from Project Team Members of enterprises from different sectors to question 14 of the research survey.

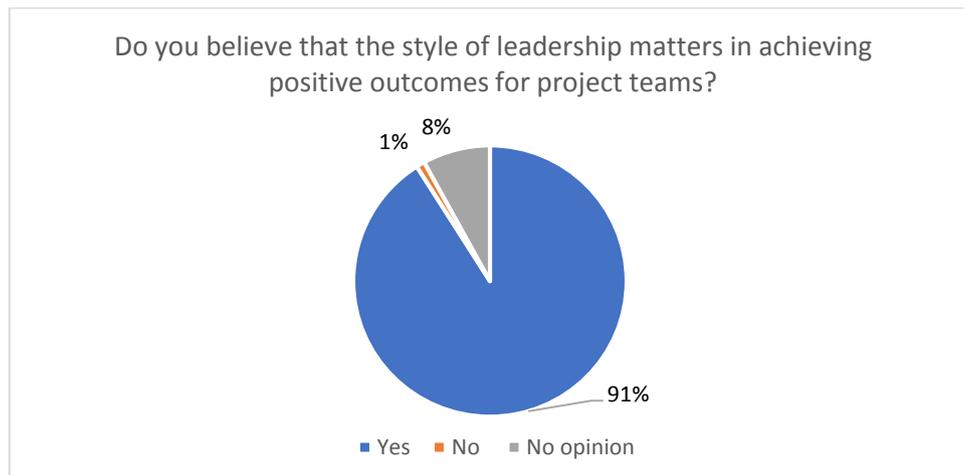

**Figure 6.** Answers from Project Team Members of enterprises from different sectors to question 15 of the research survey.

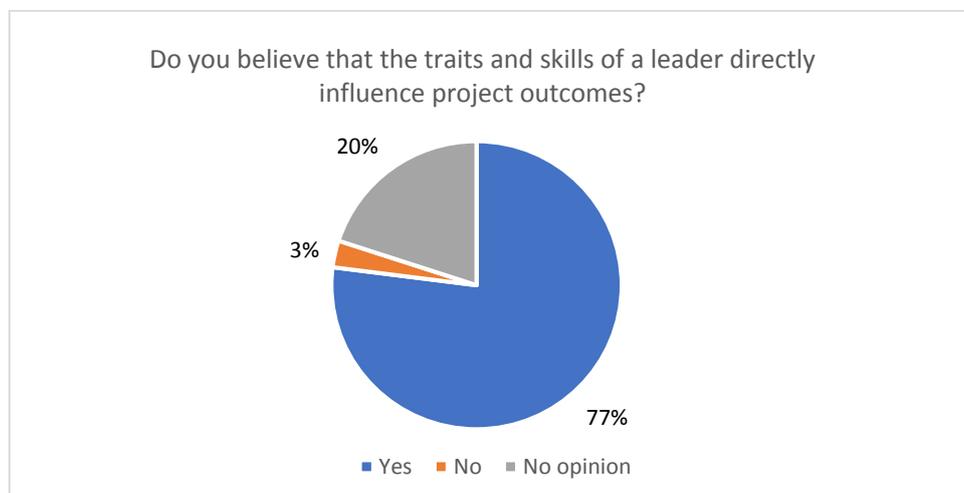

**Figure 7.** Answers from Project Team Members of enterprises of different sectors to question 16 of the research survey.

Data from 100 respondents indicate that leadership behaviors such as providing constructive feedback, clearly defining roles, encouraging independent problem-solving, and effectively communicating project goals significantly contribute to project success. The statistical analysis identified constructive feedback as the most strongly correlated factor with achieving project objectives. However, the study also revealed areas needing improvement, particularly in time management, conflict resolution, and engaging team members in decision-making processes. Given the limited sample size, further research with broader demographic and industry diversity is recommended to enhance the generalizability of these findings.

### 4.2. Analysis of research results

The first part of the questions focused on the leadership style of the leader. The questions addressed specific leader actions characteristic of selected leadership styles. The statistical analysis in this case includes the percentage distribution of responses as well as the calculation of means and medians. The means and medians were calculated based on the assumption that the response "Yes" = 1, the response "Varies" = 0.5, and "No" = 0.

To the question, "Does the leader effectively communicate the goals and outcomes of the project?", the majority of respondents (84%) answered "Yes". 12% of respondents chose the answer "Varies", while the remaining 4% selected "No". The mean was 0.900, and the median was 1.0.

For the question, "Does the leader manage conflicts arising in the team?", 25% of respondents answered "Yes". Most respondents (69%) chose "Varies", and the remaining 6% indicated "No". The mean was 0.595, and the median was 0.5.

The responses to the question, "Does the leader involve the team in the decision-making process?", were as follows: 36% of respondents answered "Yes", 41% chose "Varies", and the remaining 23% selected "No". The mean was 0.565, and the median was 0.5.

To the question, "Does the leader adequately delegate tasks and responsibilities?", 57% of respondents answered "Yes". 39% of respondents selected "Varies", 2% chose "No", and 2% of respondents did not answer this question. The mean was 0.781, and the median was 1.0.

The majority of respondents (86%) answered affirmatively to the question, "Does the leader ensure that team members understand their roles and responsibilities?" 11% of respondents chose "Varies," and 3% selected "No". The mean was 0.915, and the median was 1.0.

To the question, "Does the leader effectively manage time in the project?", 55% of respondents answered "Yes". 35% of respondents stated that the leader manages time "Varies", and the remaining 10% indicated that the leader "No" does not manage time effectively. The mean was 0.725, and the median was 1.0.

To the question, "Does the leader encourage the team to take initiative and solve problems independently?", the majority of respondents (85%) answered "Yes". 12% of respondents selected "Varies", while 2% chose "No". 1% of respondents did not answer this question. The mean was 0.919, and the median was 1.0.

The responses to the question, "Does the leader adapt their management style depending on the team's needs and situation?", were as follows: 45% of respondents answered "Yes", 49% chose "Varies", and 6% indicated "No", The mean was 0.695, and the median was 0.5.

In response to the question, "Does the leader provide constructive feedback?", 87% of respondents answered "Yes," 9% selected "Varies," and 4% chose "No." The mean was 0.915, and the median was 1.0.

To the question, "Does the leader inspire the team to achieve ambitious goals?", only 24% of respondents answered affirmatively. The majority of respondents (70%) chose "Varies", and 6% selected "No". The mean was 0.590, and the median was 0.5.

In summary, in this part of the survey, respondents most frequently selected "Yes" when asked about:
- providing constructive feedback by the leader (87%),
- ensuring that all team members understand their roles and responsibilities (86%),
- encouraging initiative and independent problem-solving (85%),
- effectively communicating the goals and priorities of the project (84%).

The answer "Varies" was most often chosen when the question concerned:
- inspiring the team to achieve ambitious goals (70%),
- managing conflicts by the leader (69%).

The highest number of "No" responses was given to the question about whether the leader involves the team in decision-making (23%).

An in-depth analysis, including the calculation of means and medians for the obtained responses based on the previously described assumptions, allowed for ranking various leadership elements according to respondents' evaluations. The results are as follows:
- Encouraging the team to take initiative in independently solving problems (mean: 0.919, median: 1),
- Providing constructive feedback (mean: 0.915, median: 1),
- Ensuring that team members understand their roles and responsibilities (mean: 0.915, median: 1),
- Effectively communicating project goals and priorities (mean: 0.900, median: 1),
- Adequately delegating tasks and responsibilities (mean: 0.781, median: 1),
- Effectively managing time (mean: 0.725, median: 1),
- Adapting management to the team's and project's needs (mean: 0.695, median: 0.5),
- Managing conflicts within the team (mean: 0.595, median: 0.5),
- Inspiring the team to achieve ambitious goals (mean: 0.590, median: 0.5),
- Involving the team in decision-making (mean: 0.565, median: 0.5).

Figure 8 presents the means and medians calculated for the questions included in the first section of the study.

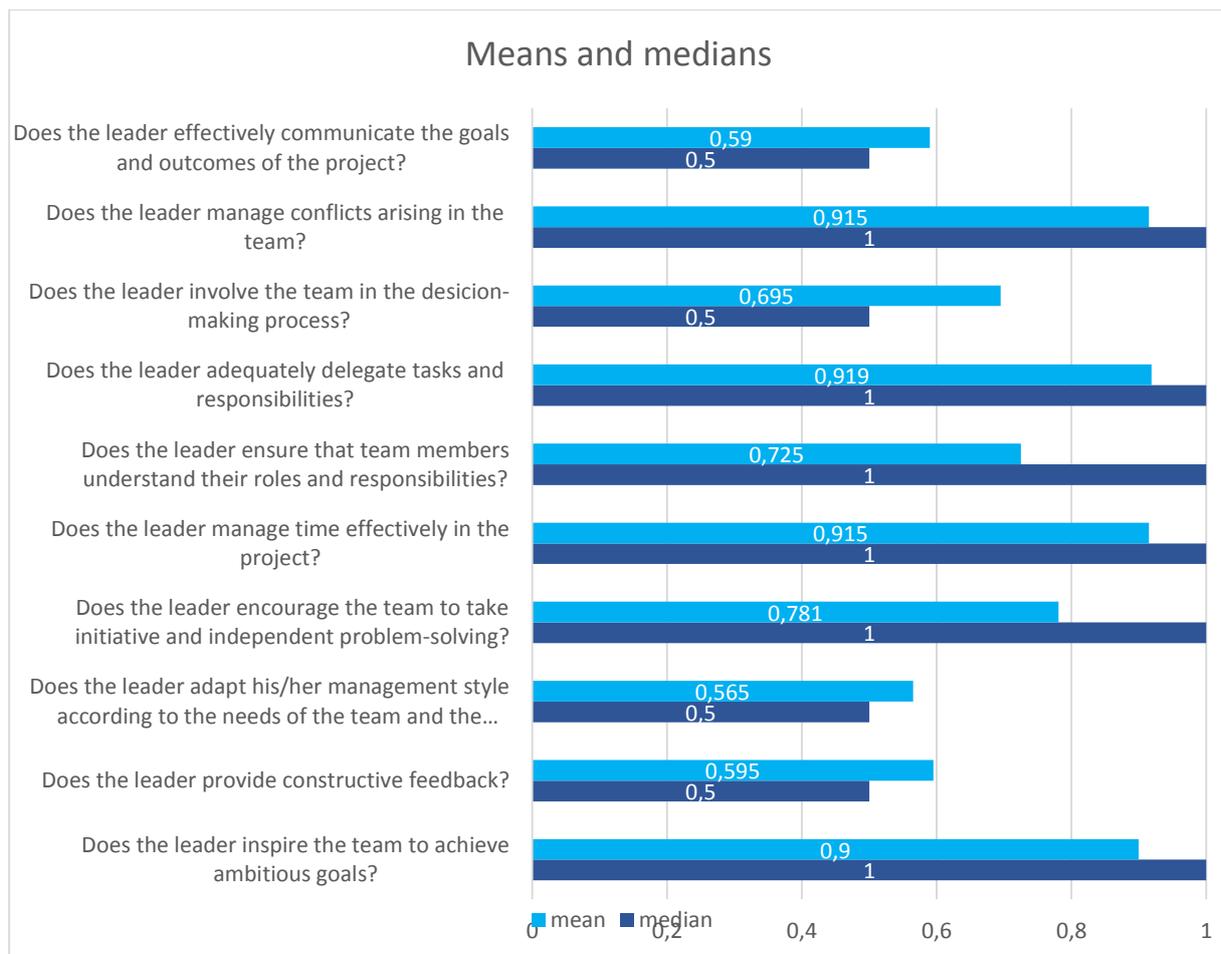

**Figure 8.** Presents the means and medians calculated for the questions included in the first section of the study.

For question 11 "Do the projects you participate in usually end successfully and meet stakeholder expectations?", the vast majority of respondents (97%) answered "Yes", while only 3% responded "No". Identical results were obtained for question 12, "Does your team usually achieve the intended project goals?" 97% of respondents answered "Yes", and 3% answered "No". For question 13, "Do projects led by your leader usually stay within the planned budget?" as many as 99% of respondents answered "Yes", with the remaining 1% responding "No". In summary, the response patterns for these three questions were quite similar.

For question 14, the vast majority of respondents, as many as 92%, answered "Yes". 2% of respondents selected "No", while the remaining 6% chose "No opinion". Similarly, for question 15, the majority of respondents (91%) answered "Yes". 1% of respondents answered "No". and 8% selected "No opinion". For the final question, the response structure was not as uniform as in the previous two questions. 77% of respondents answered "Yes", 3% responded "No", and 20% chose "No opinion".

The Spearman rank correlation analysis was conducted to examine the relationship between leaders providing constructive feedback and teams achieving their project goals. The analysis used survey responses to the questions, "Does your leader provide constructive feedback?" and "Does your team usually achieve its intended project goals?".

The results yielded a Spearman correlation coefficient (ρ) of 0.286, indicating a weak positive correlation. This suggests that teams with leaders who provide constructive feedback are more likely to achieve their project goals. To assess the statistical significance of this correlation, a significance test for the Spearman coefficient was performed. The resulting p-value was 0.004, which is less than the 0.05 threshold for significance.

This indicates that the correlation is statistically significant at the 5% level, meaning the relationship between the variables is unlikely to be random and can be considered reliable.

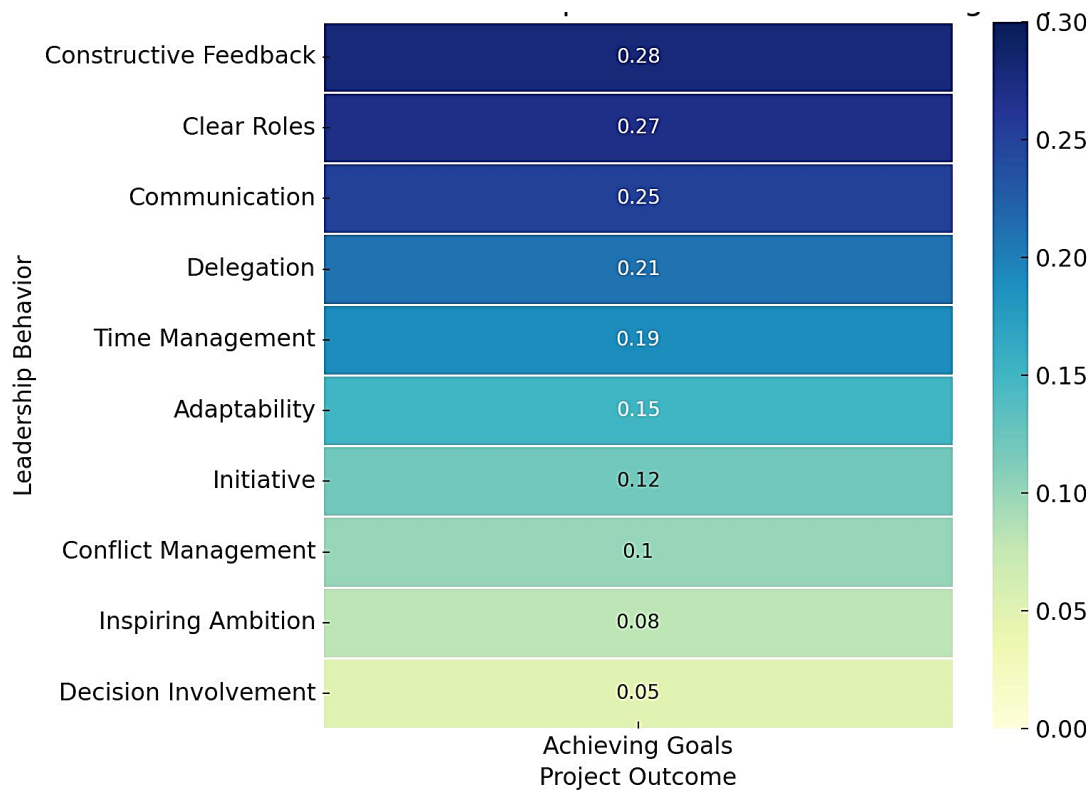

**Figure 9.** Presents the Correlation Between Leadership Behaviors and Achieving Project Goals.

This graph visualizes the strength of correlation between selected leadership behaviors and the likelihood of achieving project goals, based on survey responses from project team members. The values displayed represent Spearman correlation coefficients, where values closer to 1 indicate a stronger positive relationship.

Strongest Positive Correlations:
- Providing Constructive Feedback (ρ = 0.28): This behavior shows the highest correlation with project success, confirming that teams with leaders who offer regular, constructive input are more likely to reach project goals.
- Clearly Defining Roles (ρ = 0.27) and Effectively Communicating Goals (ρ = 0.25) also demonstrate strong links to performance, emphasizing the importance of clarity and direction in leadership.

Moderate Correlations:

- Delegating Responsibilities ($\rho = 0.21$) and Time Management ($\rho = 0.19$) indicate that operational efficiency contributes meaningfully to outcomes but may be less impactful than interpersonal leadership behaviors.
- Adaptability ($\rho = 0.15$) and Encouraging Initiative ($\rho = 0.12$) suggest that flexible and empowering leadership has moderate value in driving success.

Weaker Correlations:

- Conflict Management ($\rho = 0.10$) and Inspiring Ambitious Goals ($\rho = 0.08$) showed lower correlation levels, possibly due to varied individual perceptions or inconsistent application.
- Involving Team in Decision-Making ($\rho = 0.05$) was the weakest, which could suggest that while democratic leadership is valued, it might not directly translate into measurable performance outcomes in the short term.

The results suggest that clarity, communication, and feedback are the most impactful leadership behaviors in project settings. While inspiration and participation are important, their direct correlation with project success appears less immediate or more context-dependent. Leaders aiming to improve team performance should prioritize feedback, structured communication, and clear role definition.

## 5. Conclusions

The findings of this study confirm that leadership has a significant impact on project success, particularly through clear communication, constructive feedback, and role clarity. However, to further enhance project outcomes, leaders should focus on improving time management, team engagement in decision-making, and conflict resolution strategies.

The study confirmed that leadership significantly influences project efficiency, with 92% of respondents recognizing its direct impact on project outcomes. Similarly, 91% agreed that leadership style plays a vital role in achieving project success. These findings align with the conclusions drawn by Bwalya (2023), who emphasized that leadership style shapes team motivation, communication, and overall effectiveness. Likewise, Shrivastava and Mathur (2025) confirmed a strong relationship between leadership style and employee engagement, which indirectly boosts project performance.

Our research adds empirical support to these claims by identifying specific leadership behaviors that contribute most to project efficiency:
- Encouraging team initiative - leaders who motivate team members to independently solve problems were highly rated.
- Providing constructive feedback - effective feedback was strongly correlated with achieving project goals.
- Ensuring role clarity - leaders who ensure that team members understand their responsibilities achieved satisfactory project outcomes as well.
- Communicating project goals effectively - clear communication of goals and priorities emerged as a critical factor.

While leadership attributes like communication and constructive feedback were highly rated, several areas for improvement were noted such as:
- Time management.
- Engaging team members in decision process.
- Conflict management.

Compared to previous research, which focused primarily on general leadership outcomes, our study provides a more project-specific perspective. It bridges the gap between abstract leadership theory and concrete project management practices by offering measurable insights into behaviors that drive efficiency.

This study had several limitations. The sample was limited to 100 respondents and did not assess self-perceived leadership styles, which would provide more nuanced insights. Future studies should explore:
- Broader, cross-industry samples including varied cultural and organizational settings.
- Longitudinal research to measure the impact of leadership behaviors on project success over time.

The findings of this study confirm that leadership has a significant impact on project success, particularly through clear communication, constructive feedback, and role clarity. However, to further enhance project outcomes, leaders should focus on improving time management, team engagement in decision-making, and conflict resolution strategies. The following practical implications would be recommended:
- Develop structured training programs focused on constructive feedback, communication, and conflict management.
- Promote leadership flexibility by introducing models such as situational leadership, which proved effective in varying project contexts.
- Encourage participative practices to involve team members in decision-making, which fosters engagement and ownership.
- Implement tools for leadership self-assessment, allowing leaders to identify their dominant style and their alignment with team needs.

In conclusion, our study contributes to the growing body of knowledge on leadership in project environments by offering both empirical validation and practical guidance. Effective leadership - defined by communication as well clarity - remains a cornerstone of project success. However, addressing identified weaknesses and pursuing further research will be essential to equip future leaders with the tools necessary for navigating increasingly complex project landscapes.

## References

1. Avolio, B.J., Bass, B.M. (2004). *Multifactor Leadership Questionnaire.* Redwood City: Mind Garden.
2. Bass, B.M., Avolio, B.J. (1994). *Improving Organizational Effectiveness Through Transformational Leadership.* Thousand Oaks: Sage Publications.
3. Bass, B.M., Riggio, R.E. (2006). *Transformational Leadership.* New York: Psychology Press.
4. Blanchard, K. (2007). *Leading at a Higher Level: Blanchard on Leadership and Creating High Performing Organizations.* Upper Saddle River: FT Press.
5. Blanchard, K., Zigarmi, D., Nelson, R. (1993). Situational Leadership® after 25 years: A retrospective. *Journal of Leadership & Organizational Studies, vol. 1, no. 1.*
6. Burns, J.M. (1978). *Leadership.* New York: Harper & Row.
7. Bwalya, A.R. (2023). Leadership Styles. *Journal of Entrepreneurship, Management and Innovation, vol. 11, no. 8.*
8. Chmiel, N. (2003). *Psychologia pracy i organizacji.* Gdańsk: Gdańskie Wydawnictwo Psychologiczne.
9. Gastil, J. (1994). A definition and illustration of democratic leadership. *Human Relations, vol. 47, no. 8.*
10. Greenleaf, R.K. (1977). *Servant Leadership: A Journey into the Nature of Legitimate Power and Greatness.* New York: Paulist Press.
11. Kotarbiński, T. (2000). *Traktat o dobrej robocie.* Wrocław: Ossolineum.
12. Koźmiński, A.K., Jemielniak, D. (2011). *Zarządzanie. Teoria i praktyka.* Warszawa: PWN.
13. Northouse, P.G. (2016). *Leadership: Theory and Practice.* Thousand Oaks: Sage Publications.
14. Obłój, K. (2007). *Strategia organizacji.* Warszawa: PWE.
15. Rubin, K. (2012). *Essential Scrum: A Practical Guide to the Most Popular Agile Process.* Upper Saddle River: Addison-Wesley.
16. Schultz, D.P., Schultz, S.E. (2006). *Psychologia a wyzwania dzisiejszej pracy.* Warszawa: PWN.

17. Shrivastava, A., Mathur, S. (2025). *The Relationship between Leadership Style and Employee Engagement*.
18. Spears, L.C. (2010). *Practicing Servant Leadership: Succeeding Through Trust, Bravery, and Forgiveness*. San Francisco: Jossey-Bass.
19. Stocki, R., Prokopowicz, P., Żmuda, G. (2008). *Pełna partycypacja w zarządzaniu*. Kraków: Wolters Kluwer.
20. Wlizło, I. (2021). Zwinne metodyki zarządzania projektami w teorii i praktyce. *Zeszyty Naukowe Wyższej Szkoły Zarządzania Ochroną Pracy w Katowicach, vol. 1, no. 17*.
21. Yukl, G.A. (2013). *Leadership in Organizations*. Boston: Pearson.
22. Zaccaro, S.J. (2007). Trait-based perspectives of leadership. *American Psychologist, vol. 62, no. 1*.